\documentclass[aps,prl,twocolumn,showpacs,superscriptaddress]{revtex4-1}
\usepackage{amsmath}
\usepackage{epsfig}
\usepackage{color}
\usepackage{endnotes}
\usepackage{natbib}
\usepackage[colorlinks,breaklinks]{hyperref}
\usepackage{nicefrac}
\usepackage{bm}
\usepackage{dcolumn}
\usepackage{graphics}
\usepackage{graphicx} 
\usepackage{tikz} 
\usepackage{amsmath}
\usepackage{amssymb} 
\usepackage{color}
\usepackage{changes}
\usepackage{multirow}
\usepackage{lineno}
\usepackage[normalem]{ulem}
\usepackage{xcolor}

\newcommand{\be}{\begin{equation}}
\newcommand{\ee}{\end{equation}}
\newcommand{\bea}{\begin{align}}
\newcommand{\eea}{\end{align}}
\newcommand{\beqa}{\begin{eqnarray}}
\newcommand{\eeqa}{\end{eqnarray}}

\newcommand*{\MIT }{Massachusetts Institute of Technology, Cambridge, Massachusetts 02139, USA}
\newcommand*{\ODU}{Old Dominion University, Norfolk, Virginia 23529, USA}

\newcommand*{\TAU }{School of Physics and Astronomy, Tel Aviv University, Tel Aviv 69978, Israel}

\newcommand*{\GW}{George Washington University, Washington, D.C., 20052, USA}

\newcommand*{\liverpool}{University of Liverpool, Liverpool L69 7ZE, United Kingdom}
\newcommand*{\FNAL}{Fermi National Accelerator Laboratory, Batavia, IL 60510, USA}
\begin{document}

\title{Transport Estimations of Final State Interaction Effects on Short--range Correlation Studies Using the $(e,e'p)$ and $(e,e'pp)$ Reactions}

\author{N. Wright}
\affiliation{\MIT}
\author{A. Papadopoulou}
\affiliation{\MIT}
\author{J.R. Pybus}
\affiliation{\MIT}
\author{S. Gardiner}
\affiliation{\FNAL}
\author{M. Roda}
\affiliation{\liverpool}
\author{F.~Hauenstein}
\affiliation{\MIT}
\affiliation{\ODU}
\author{A. Ashkenazi}
\affiliation{\TAU}
\author{L. Weinstein}
\affiliation{\ODU}
\author{A. Schmidt}
\affiliation{\GW}
\author{O. Hen}
\email[Contact Author \ ]{(hen@mit.edu)}
\affiliation{\MIT}

\begin{abstract}
  Short range correlated (SRC) nucleon-nucleon pairs in nuclei are typically studied using measurements of electron-induced hard nucleon-knockout reactions (e.g. $(e,e'p)$ and $(e,e'pN)$), where the kinematics of the knocked-out nucleons are used to infer their initial state prior to the interaction.  The validity of this inference relies on our understanding of the scattering reaction, most importantly how rescattering of the detected nucleons (final state interactions or FSI) distort  their kinematical distributions.  Recent SRC measurements on medium to heavy nuclei have been performed at high-$x_B$ (i.e.,  anti-parallel kinematics) where calculations of light nuclei indicate that such distortion effects are small.  
  Here we study the impact of FSI on recent $^{12}$C$(e,e'p)$ and $^{12}$C$(e,e'pp)$ measurements using
  a transport approach.
We find that while FSI can significantly distort the measured kinematical distributions of SRC breakup events, selecting high-$x_B$ anti-parallel events strongly suppresses such distortions. In addition, including the effects of FSI improves the agreement between Generalized Contact Formalism-based calculations and data and can help identify those observables that have minimal sensitivity to FSI effects. 
This result helps confirm the interpretation of experimental data in terms of initial-state momentum distributions and provides a new tool for the study of SRCs using lepton-scattering reactions.
\end{abstract}

\maketitle


Short-range correlations (SRCs) are pairs of strongly interacting nucleons at short distances with correspondingly large relative momentum and smaller center-of-mass momentum.  Studies of SRC properties allow the investigation of the relation between the single-particle, mean-field structure of nuclei and their fully correlated many-body wavefunction~\cite{CiofidegliAtti:1995qe,ryckebusch15,colle15,Cruz-Torres:2019fum}, while providing insight into the nature of the strong nuclear interaction at short distance~\cite{subedi08,korover14,schmidt20,Korover:2020lqf} and the impact of the nuclear medium of the structure of bound nucleons~\cite{weinstein11,Hen12,Hen:2013oha,Chen:2016bde,Schmookler:2019nvf,Segarra:2019gbp}.  
See recent reviews~\cite{Hen:2016kwk,Atti:2015eda}.

Much of our experimental understanding of SRCs comes from electron-induced nucleon knockout measurements~\cite{subedi08,shneor07,korover14,Hen:2014nza,duer18,Duer:2018sxh,Cohen:2018gzh,schmidt20,Korover:2020lqf}.  Here high-energy electrons are scattered off nuclei followed by the detection of the scattered electron and one or two emitted nucleons from the breakup of an SRC pair.  
The initial state of the SRC pair in the nucleus is then inferred from the detected momenta (e.g., by subtracting the momentum transfer from the detected struck nucleon momentum to infer its initial momentum). 

This interpretation is sensitive to rescattering of the outgoing nucleons by each other or by the residual nucleus, i.e., by the effects of nuclear final-state interactions (FSI). Such effects can either distort the inferred distributions or decrease their magnitude by rescattering a nucleon so that the event falls outside the experimental acceptance.  The latter flux-attenuation effect is commonly referred to as nuclear ``transparency''~\cite{Sargsian:2001ax,Dutta:2012ii,Colle:2015lyl}.

To minimize such interpretation complications, SRC measurements are performed in specific kinematics where model calculations suggest FSI distortions are minimized and transparency effects are well understood. 
These calculations usually employ Glauber theory in an Eikonal approximation and are reliable for light nuclei at large enough nucleon momenta (i.e. $A=2,3$)~\cite{Frankfurt:1996xx,Sargsian:2001ax,frankfurt08b,Colle:2015lyl}. 

For medium and heavy nuclei however full calculations are not feasible and approximate calculations are less reliable, especially for estimating kinematical distortion effects.
In addition, the Glauber approximation is only valid for outgoing nucleons with energy higher than a few hundred MeV
which applies for the struck nucleon in large momentum transfer $(e,e'p)$ and $(e,e'pN)$ reactions but not for the recoil nucleon from an SRC pair measured in $(e,e'pN)$ reactions.

Therefore, the conclusion that FSI distortion effects are small in the selected kinematics is based on (a) extrapolating light-nuclei calculations to heavier nuclei, (b) the localization of the high-energy interaction and (c) the observed agreement between experimentally-extracted momentum distributions and Plane Wave Impulse Approximation (PWIA) calculations that include transparency effects but not FSI distortions~\cite{Cohen:2018gzh,schmidt20,Pybus:2020itv,Korover:2020lqf}.
It is thus beneficial to study the impact of FSI on SRC measurements using complementary theoretical approaches that are suitable for medium and heavy nuclei studies.

One such approach uses effective transport theory, where effective nucleon-nucleon scattering cross-sections are used to calculate the propagation of nucleons as they traverse the nuclear medium.
Here we use such calculations to study the effects of FSI on semi-exclusive $^{12}$C$(e,e'p)$ and exclusive $^{12}$C$(e,e'pp)$ SRC measurements.
We assess the effectiveness of the kinematical cuts applied in previous SRC studies in suppressing FSI distortions and hence increasing the sensitivity to initial-state nuclear distributions. We specifically focused on the measurements of Refs.~\cite{schmidt20,Cohen:2018gzh} which used the CLAS spectrometer~\cite{Mecking:2003zu} with a $5.01$ GeV electron beam at $x_B \ge 1.2$, $Q^2 \gtrsim 1.7$ GeV$^2$, and large $(e,e'p)$ missing-momentum (where $x_B = Q^2/2m\omega$, $Q^2=\mathbf{q}^2 -\omega^2$, $m$ is the nucleon mass, and $\mathbf{q}$ and $\omega$ are the three-momentum and energy transfers, respectively).

Our calculations used the Generalized Contact Formalism (GCF) to model the ground-state distribution of SRC pairs in nuclei~\cite{Weiss:2015mba,Weiss:2016obx,Cruz-Torres:2019fum}, in conjunction with the electron-scattering version of the GENIE lepton-nucleus scattering event generator~\cite{Andreopoulos:2009rq,Papadopolou:2020zkd}.
The latter simulates the scattering process and accounts for FSI via a data-driven IntraNuclear Cascade transport model~\cite{Andreopoulos:2009rq,Andreopoulos:2015wxa,Dytman:2011zz,Dytman:2021ohr}.  GENIE is used extensively to simulate neutrino-nucleus interactions, e.g.~\cite{Fiorentini13b,Agafonova:2015jxn,Adamson:2017qqn,Adamson:2017gxd,Abratenko:2020acr}.
Its electron-scattering version, $e$GENIE~\cite{Papadopolou:2020zkd}, was devised to use electron-nucleus scattering data to help constrain vector-interaction aspects of neutrino-nucleus scattering.  This is the first use of $e$GENIE to study the effects of FSI in electron-scattering data.

eGENIE generates quasielastic (QE) scattering events according to the PWIA cross section, which is the product of the electron-nucleon cross section and the probability to find a nucleon in the nucleus with a given initial energy and momentum, as defined by the nuclear ground state spectral function (or by the momentum distribution for Fermi gas models). 

Current $e$GENIE models typically use  a Fermi gas distribution (global or local) with a fixed binding energy and thus do not directly account for SRCs.
Some models do include the single-nucleon aspects of SRCs by including a high-momentum tail to the single-nucleon momentum distribution above the Fermi momentum, either empirically following the Bodek-Ritchie formulation or via the use of more modern spectral functions~\cite{Bodek:1980ar,Bodek:1981wr,Bodek:2014pka}.

In order to fully describe SRC pairs in $e$GENIE via two-body densities, we implemented the GCF model.  The GCF model provides a pair-wise approximation for the nuclear ground-state spectral function at short-distance and high-momentum.
This accounts for both the ``struck'' nucleon (which absorbs the momentum transfer), as well as its correlated partner nucleon that is also emitted in the breakup of an SRC pair.
The combined treatment of both partners enables modeling of electron scattering from SRC pairs via QE two-nucleon knockout  $(e,e'NN)$ reactions.

In the GCF, the proton (or neutron) spectral function is approximated by a sum over all possible $NN$-SRC pairs~\cite{Weiss:2016obx,Weiss:2018tbu}:
\begin{equation}
\begin{split}
S^{p(n)}(\epsilon_1,p_1) =\;\;  &C_{pn}^{s=1}S_{pn}^{s=1}(\epsilon_1,p_1) + C_{pn}^{s=0}S_{pn}^{s=0}(\epsilon_1,p_1) \\
+ 2& \cdot C_{pp(nn)}^{s=0}S_{pp(nn)}^{s=0}(\epsilon_1,p_1),
\end{split}
\end{equation}
where $S^{p(n)}(E,k)$ is the probability of finding a proton (or neutron) with momentum $p_1$ and off-shell energy $\epsilon_1$. $C_{NN}^{s=i}$ are contact coefficients defining the abundance of SRC pairs of isospin $NN$ (= $np, pp, nn$) and total spin $i$ (= $0,1$). $S_{NN}^{s=i}(\epsilon_1,p_1)$ is the SRC pair spectral function given by:
\begin{equation}
\begin{split}
S_{NN}^{s=i}(\epsilon_1,p_1) = \frac{1}{4\pi} \int \frac{d{\bf p_2}}{(2\pi)^3} &\delta\left(f(p_2)\right) \left| \psi_{NN}^{s=i}\left(\frac{\bf{p}_1-\bf{p}_2}{2} \right)\right|^2 \\ 
&n_{NN}^{s=i}(\bf{p}_1+\bf{p}_2),
\end{split}
\label{eq:2}
\end{equation}
where $p_2$ is the momentum of the recoil nucleon from the SRC pair, $\psi_{NN}^{s=i}$ is the two-nucleon wave-function calculated from the appropriate $NN$ potential, $\vert\psi_{NN}^{s=i}\vert^2$ and $n_{NN}^{s=i}$ are the respectively relative and center-of-mass (c.m.) momentum distributions of SRC pairs, and $\delta\left(f(p_2)\right)$ is an energy conserving delta-function.  See Ref.~\cite{Weiss:2018tbu} for details.

The GCF successfully reproduces both ab-initio calculations of one- and two-nucleon densities at both high-momenta ($p\ge 300$ MeV/c) and short-distance~\cite{Weiss:2016obx,Cruz-Torres:2019fum}, as well as measured electron and proton scattering data~\cite{Duer:2018sxh,schmidt20,Pybus:2020itv,Korover:2020lqf}.

\begin{figure}[t]
\begin{center}
\includegraphics[height=4.5 cm, width=0.9\columnwidth]{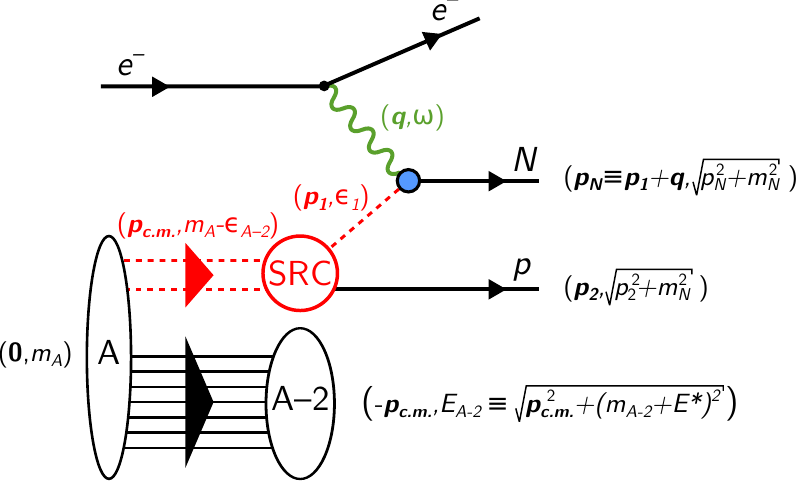}
	\caption{
	Reaction diagram for quasielastic (QE) electron scattering off nucleons in an SRC pair. The nuclear ground state is assumed to be factorized according to the GCF into an off-shell SRC pair with momentum ${\bf{p}}_{c.m.}$ and a spectator $A-2$ nuclear system with momentum $-{\bf{p}}_{c.m.}$. In the limit of large momentum-transfer, the nucleons in the SRC pair are assumed to further factorize into an active off-shell (lead) nucleon that absorbs the momentum transfer and an on-shell recoil nucleon that is a spectator to the reaction.
	}
\label{Fig:diagram}
\end{center}
\end{figure}

We modeled QE scattering from SRC pairs using the GCF by assuming that the electron scatters off a single nucleon via the exchange of a virtual photon with momentum $\bf{q}$ and energy $\omega$. The struck nucleon has momentum $\bf{p}_1$ and energy $\epsilon_1$ before the interaction, and momentum ${\bf{p}}_1 + {\bf{q}}$ and energy $\epsilon_1+\omega = \sqrt{({\bf{p}}_1 + {\bf{q}})^2+m^2}$ right after  (see Fig.~\ref{Fig:diagram}).
Knocking out one nucleon from an SRC pair will result in the emission of a correlated recoil nucleon, which is treated as an on-shell spectator.
Energy and momenta are conserved for each event based on the diagram shown in Fig.~\ref{Fig:diagram}.  
In the PWIA (i.e., without FSI), the nucleons will exit the nucleus with these momenta. With FSI they can re-scatter, in which case their kinematics will be modified.

The PWIA cross-section can be factorized as $\sigma_{(e,e'p)}^{PWIA} = K \sigma_{ep} S^p(\epsilon_1,p_1)$ where $K$ is a kinematical factor and $\sigma_{ep}$ is an off-shell electron-nucleon interaction cross-section. The $(e,e'pp)$ PWIA cross-section has an equivalent factorized form that is achieved by omitting the integration over $\bf{p}_2$ in Eq.~\ref{eq:2}. See Ref.~\cite{Pybus:2020itv} for details.

Previous studies corrected their PWIA calculations for selected FSI effects: outgoing nucleon attenuation using nuclear transparency (T) factors and single-charge exchange (SCX) (e.g., $(n,p)$ and $(p,n)$ reactions where a struck proton emerged as a neutron and vice versa)~\cite{Duer:2018sxh, schmidt20,Korover:2020lqf}.  They did not correct for FSI-related distortions.  The corrected ``T+SCX'' cross sections are calculated as follows:
\begin{equation} 
\begin{split}
\sigma^{T+SCX}_{A(e,e'p)} = &\sigma^{PWIA}_{A(e,e'p)} \cdot P_{A}^{p}\cdot T_{A,p} + \\
& \sigma^{PWIA}_{A(e,e'n)}\cdot P_{A}^{[n]}\cdot T_{A,p}, \\ \\
\sigma^{T+SCX}_{A(e,e'pp)} = &\sigma^{PWIA}_{A(e,e'pp)} \cdot P_{A}^{pp}\cdot T_{A,pp} + \\
& \sigma^{PWIA}_{A(e,e'np)}\cdot P_{A}^{[n]p}\cdot T_{A,pp} + \\
& \sigma^{PWIA}_{A(e,e'pn)}\cdot P_{A}^{p[n]}\cdot T_{A,pp}, \\ \\
\end{split}
\label{eq:scx_gcf}
\end{equation}
where $T_{A,p}$ and $T_{A,pp}$ are the nuclear transparency factors for one- and two-proton knockout.
$P^{NN}_{A}$ and $P^{N}_{A}$ are the probabilities for one- or two nucleon knockout without subsequent SCX,
and $P^{[N]}_{A}$, $P^{[N]N}_{A}$ and $P^{N[N]}_{A}$ are the probabilities for
the nucleon in brackets to undergo SCX, respectively.  The first or leading nucleon is the struck nucleon and the second is the spectator recoil nucleon.

The values of these parameters are detailed in Ref.~\cite{Duer:2018sxh} (also used by Refs.~\cite{schmidt20,Korover:2020lqf}), based on the calculation of Ref.~\cite{Colle:2015lyl}.  
They depend on the kinematics of the measured reaction and were studied for the kinematics of the CLAS data analyzed herein. While the calculation of Ref.~\cite{Colle:2015lyl} can account for full kinematical distortions, the use of their calculated Transparency and SCX probabilities in Eq.~\ref{eq:scx_gcf} is effective and only accounts for the average effect of flux reduction and zero-angle $(n,p)$ and $(p,n)$ reactions, without considering small angle rescattering that has a stronger impact on the measured kinematical distributions.
It does however introduces the important effects of scattering off $np$-SRC pairs that are experimentally detected as $pp$-SRCs and vise-versa where the different underlaying distribution of these reactions lead to an effective kinematical distortion of the measured distributions.
For example, the yield of final (measurable) $(e,e'pp)$ events is the number of $(e,e'pp)$ events that do not undergo SCX ($P^{pp}_A$) times the transparency factor for the two protons to emerge from the nucleus ($T_{A,pp}$) plus the number of $(e,e'np)$ events times the probability that the neutron undergoes SCX to become a proton and the proton does not undergo SCX ($P^{[n]p}_A$) times the transparency  factor for the two protons to emerge from the nucleus ($T_{A,pp}$) plus the same for $(e,e'pn)$ events.

To improve on the above and include full FSI effects in a transport approximation we implemented the GCF's two-nucleon spectral function into the $e$GENIE QE interaction calculation. Our implementation assumes the SRC pair density follows the one-body charge density which is a good approximation for a relatively small nucleus like carbon~\cite{Cruz-Torres:2019fum}. We used the same GCF model parameters as Ref.~\cite{Cruz-Torres:2019fum}, namely SRC pair relative-momentum distribution calculated with the AV18 $NN$ interaction, and a gaussian SRC pair c.m. momentum distribution~\cite{CiofidegliAtti:1995qe,Cohen:2018gzh} with a gaussian width (in each cartesian direction) of $\sigma_{c.m.} = 150$ MeV/c~\cite{Cohen:2018gzh}. 
We used the default GENIE cross-section model for QE events with the Rosenbluth prescription and free-nucleon form factors~\cite{Bradford:2006yz}.

Final state interactions were simulated using the GENIE ``hA'' model, which  uses hadron-nucleus cross sections to model the overall effect of hadron rescattering~\cite{PhysRevC.23.2173,PhysRevC.28.2548,PhysRevC.61.034601,PhysRevC.14.635,Clough:1974qt,Dytman:2021ohr}.  
For each hadron involved in the initial lepton-nucleus scattering reaction GENIE looks up the energy-dependent hadron-nucleus cross section and scales it by the integrated nuclear density that the hadron traverses on its path.  If it determines that the hadron reinteracted, then it randomly selects the reaction type (charge-exchange, inelastic, etc) according to their relative cross-sections and selects the resulting particle(s) and energies.  Each outgoing particle is treated independently and interacts at most once.

We validated our GENIE PWIA-GCF implementation (without FSI) by comparing it to the original PWIA-GCF calculations of Ref.~\cite{schmidt20} for the scattering of  $5.01$ GeV electrons from nucleons in  SRC pairs in the $^{12}$C nucleus. 

We then compared three types of calculations for our full study:
\begin{enumerate}
\item `PWIA' calculations without any reaction effects,
\item `T+SCX' calculations where Transparency effects and SCX (n,p) and (p,n) reactions were accounted for 'by hand' using input from Glauber calculations as shown in Eq.~\ref{eq:scx_gcf}, and
\item `Full' calculations where the GENIE transport model is used to account for all nuclear reaction effects experienced by the knockout nucleons.
\end{enumerate}

We calculated event distributions for equivalent experimental luminosities and passed the produced events samples through a Monte Carlo simulation of the CLAS detector~\cite{Mecking:2003zu} to  only include events within the CLAS acceptance and to smear the detected momenta by the CLAS momentum resolution.

\begin{figure}[t]
\begin{center}
\includegraphics[height=5.3 cm, width=\columnwidth]{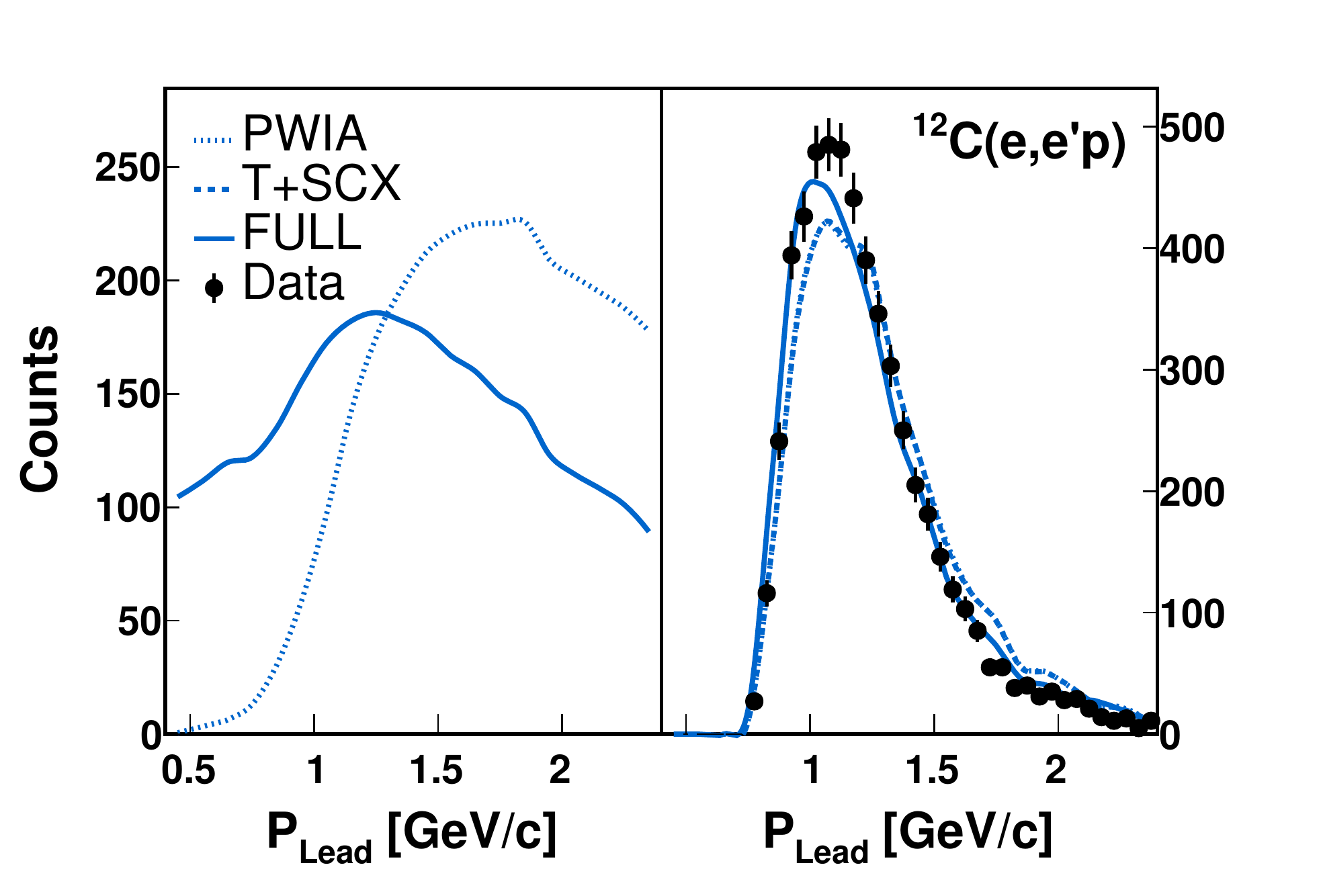}
	\caption{
	Leading-proton momentum distribution for $^{12}$C$(e,e'p)$ events for scattering off nucleons in SRC pairs as modeled by the GCF. Left panel shows calculations for all  events that pass the CLAS acceptance cuts, while the right panel shows those events that additionally pass the SRC selection cuts. The points show the data from Ref.~\cite{schmidt20}.  The dotted curve shows the PWIA, the dashed shows the T+SCX, and the solid curve shows the Full calculation.  The right panel y-axis scale correspond to the measured data counts and the calculations are individually area normalized to the data. The left panel y-axis scale is arbitrary.  Since the T+SCX calculation does not change the shape of the momentum distribution, it is almost indistinguishable from the PWIA calculation. 
	}
\label{Fig:p_lead}
\end{center}
\end{figure}

\begin{figure}[t]
\begin{center}
\includegraphics[height=5 cm, width=\columnwidth]{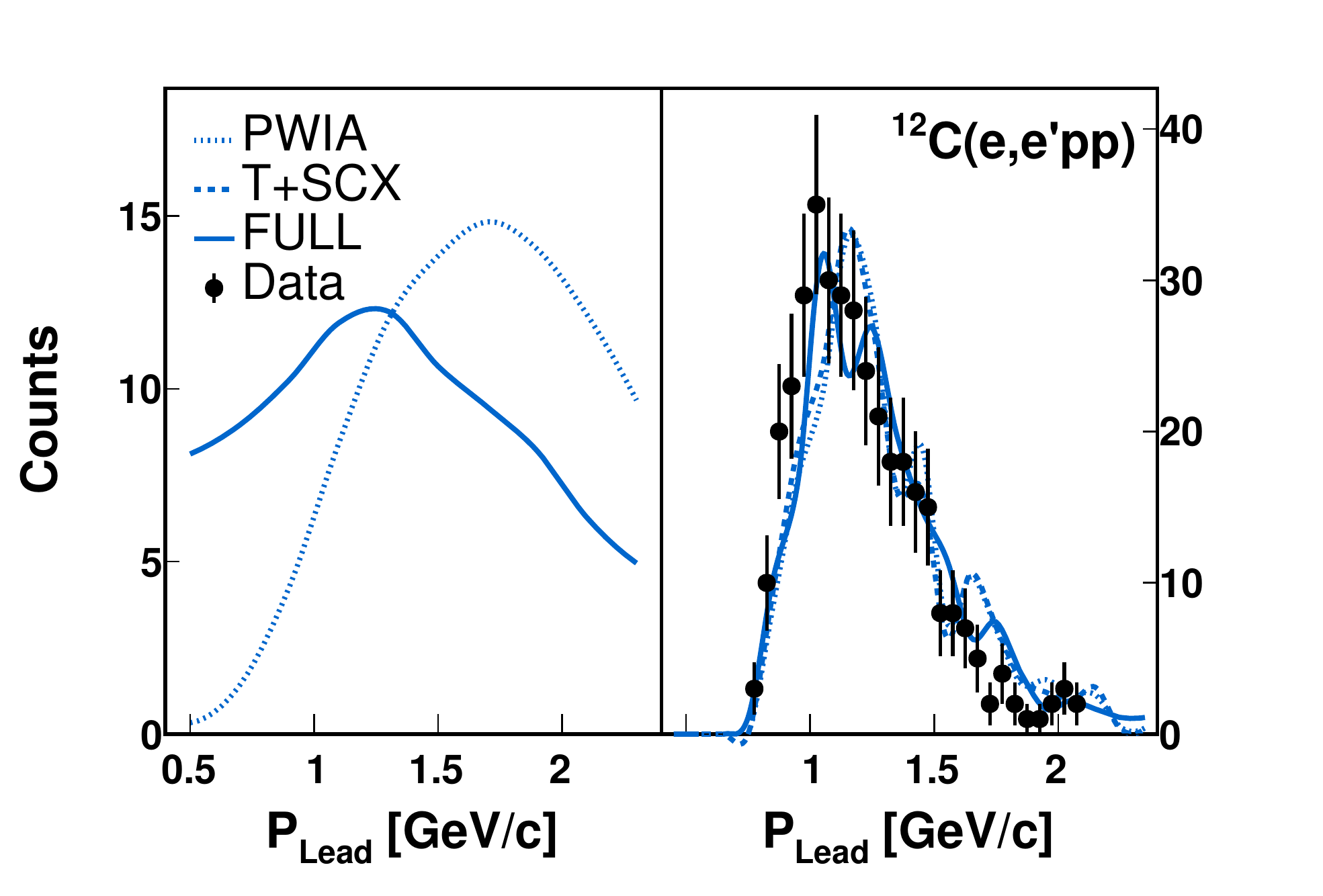}
\includegraphics[height=5 cm, width=\columnwidth]{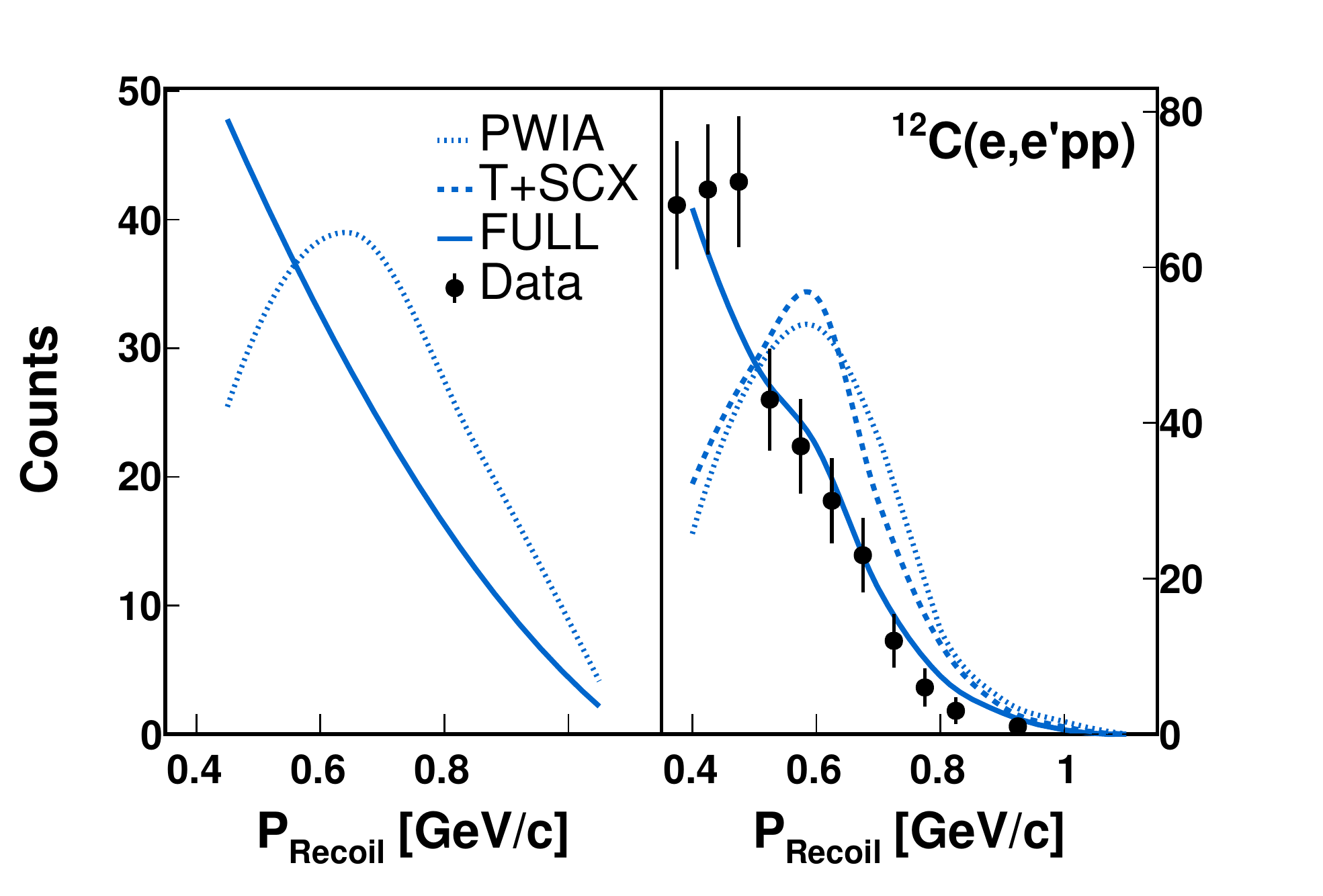}
	\caption{
	Same as Fig.~\ref{Fig:p_lead} for both the lead (top) and recoil (bottom) proton in $^{12}$C$(e,e'pp)$ reaction for scattering off nucleons in SRC pairs.
	}
\label{Fig:pp_lead_recoil}
\end{center}
\end{figure}

\begin{figure*}[t]
\begin{center}
\includegraphics[height=5.3cm, width=\columnwidth]{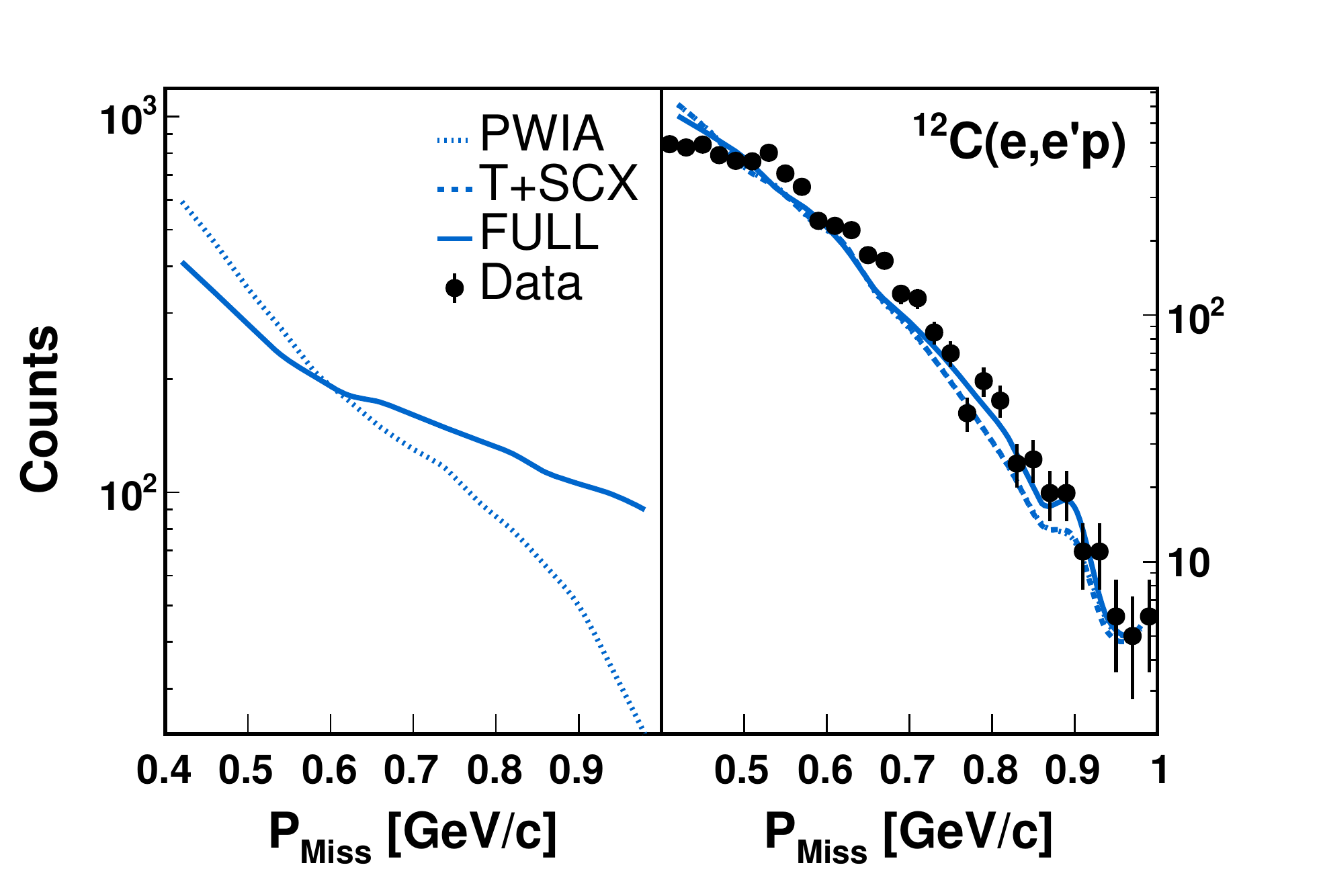}
\includegraphics[height=5.3cm, width=\columnwidth]{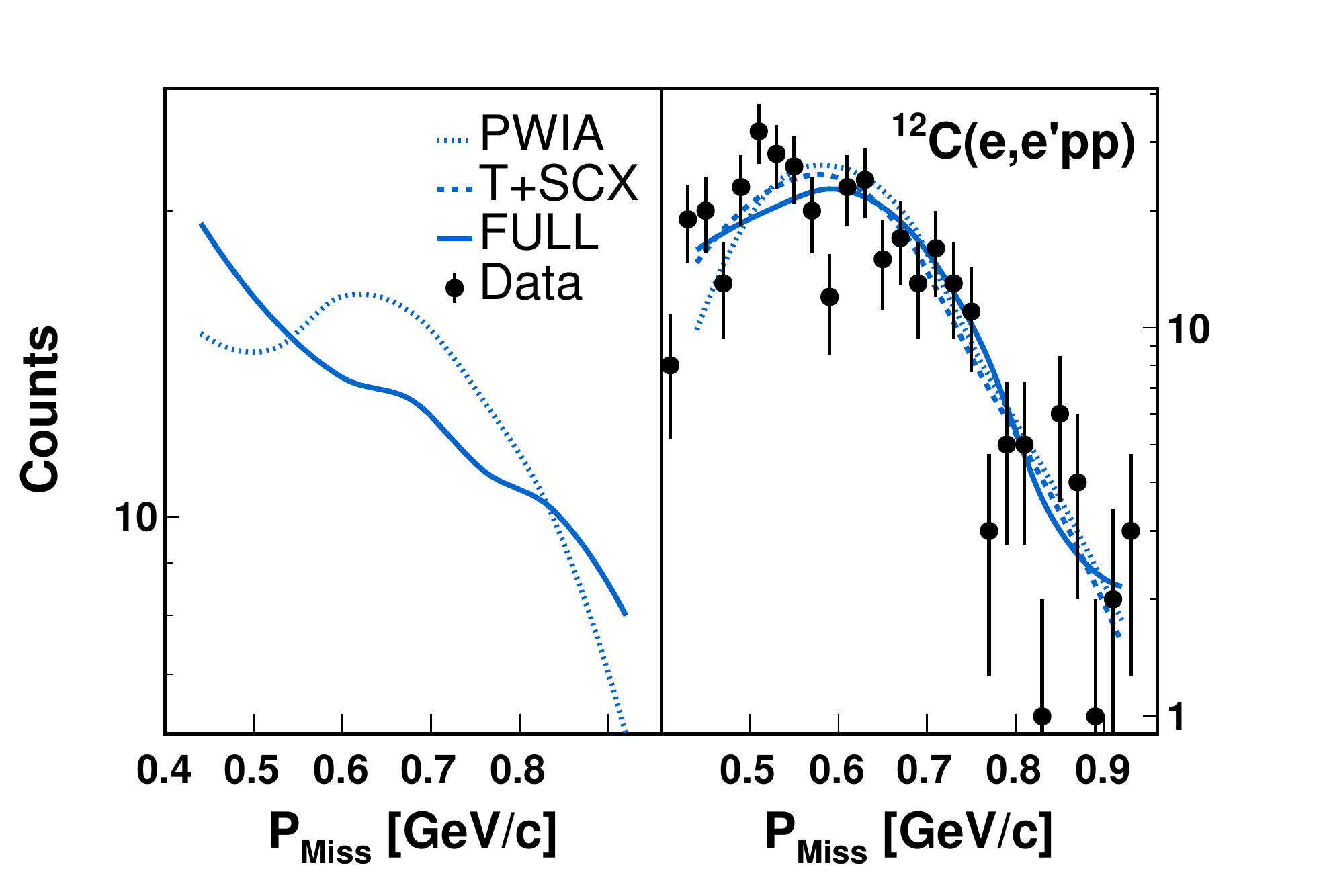}
\caption{
  Same as Fig.~\ref{Fig:p_lead}  for the leading-proton missing momentum $p_{miss} = \vert \mathbf{p}_N - \mathbf{q}\vert$ in $^{12}$C$(e,e'p)$ (left) and $^{12}$C$(e,e'pp)$ (right) events for scattering off nucleons in SRC pairs. 
}
\label{Fig:Pmiss}
\end{center}
\end{figure*}

\begin{figure}[t]
\begin{center}
\includegraphics[height=5.3 cm, width=\columnwidth]{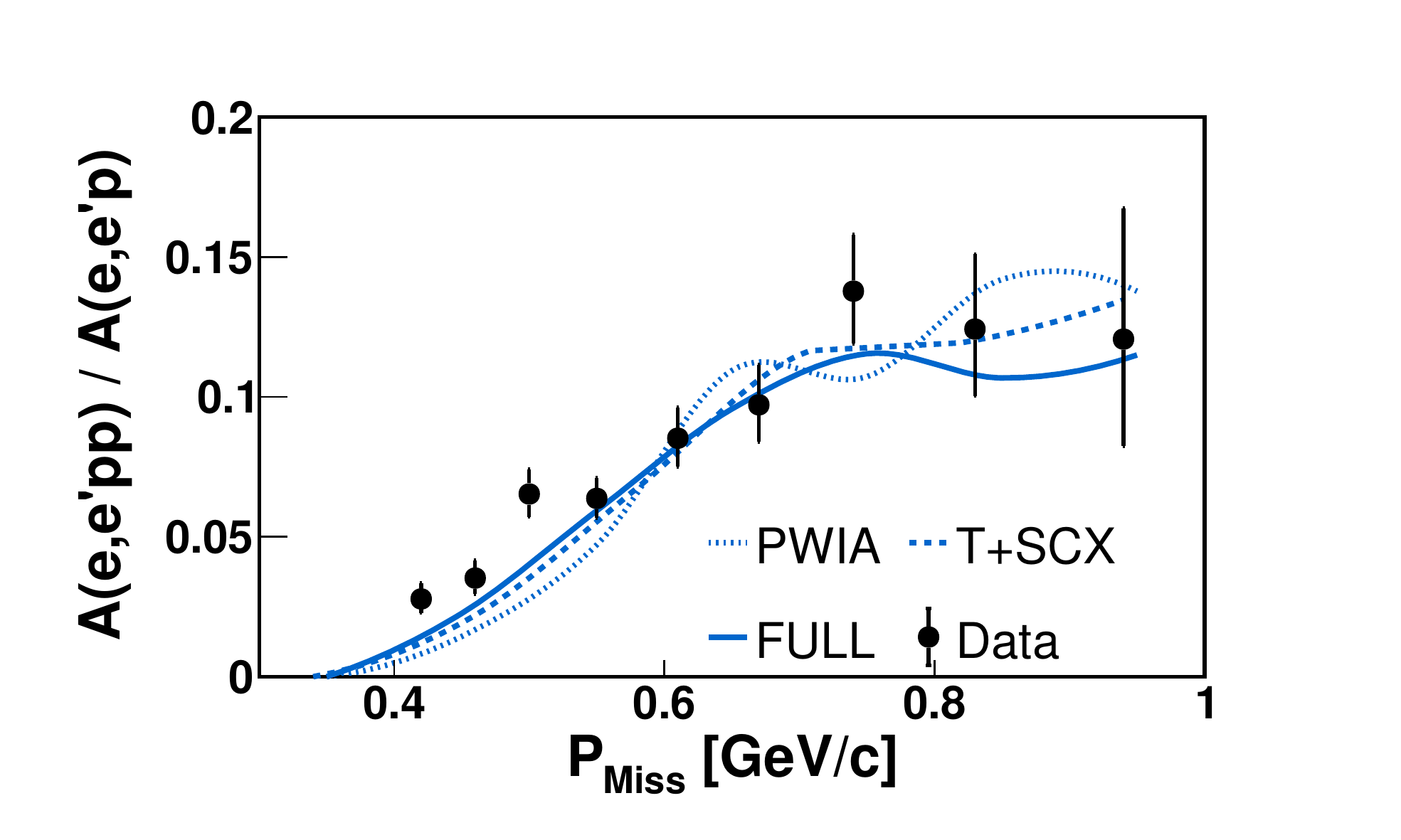}
	\caption{
	Missing momentum distribution for the $(e,e'pp)$/$(e,e'p)$ event yield ratio for $^{12}$C SRC breakup events passing the SRC event selection cuts.  The points show the data from Ref.~\cite{schmidt20}.  The dotted curve shows the PWIA, the dashed shows the T+SCX, and the solid curve shows the Full calculation.  
	}
\label{Fig:pp_p}
\end{center}
\end{figure}

\begin{figure*}[t]
\begin{center}
\includegraphics[height=7.5cm, width=1.8\columnwidth]{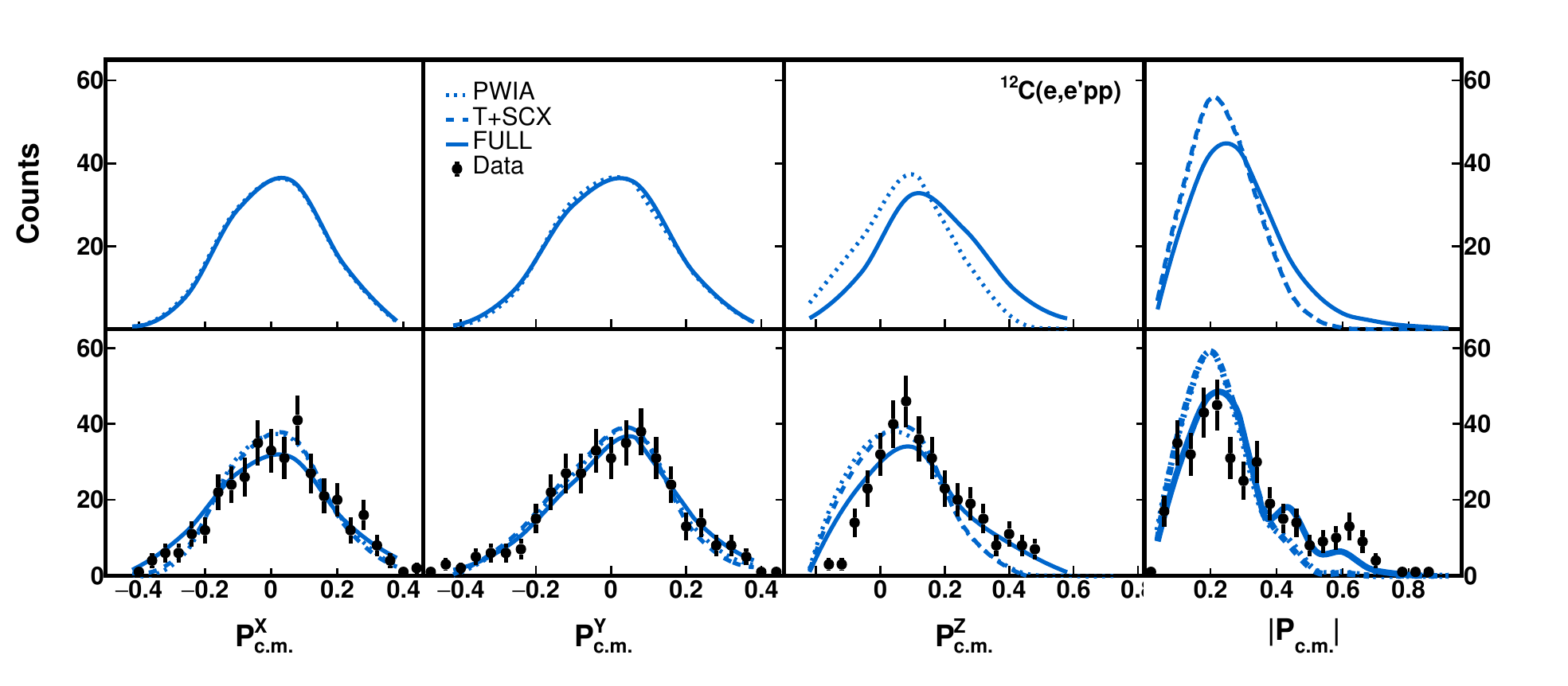}
	\caption{
          The inferred SRC pair c.m. momentum ($\mathbf{p}_{c.m.} = \mathbf{p}_{lead} + \mathbf{p}_{recoil} - \mathbf{q}$) distributions for scattering off nucleons in SRC pairs.
          The points show the data from Ref.~\cite{Cohen:2018gzh}.  The dotted curve shows the PWIA, the dashed shows the T+SCX, and the solid curve shows the Full calculation.
The top panels show calculations for all  events that pass the CLAS acceptance cuts, while the bottom panels show those events that additionally pass the SRC selection cuts. From left to right, the $x, y$ and $z$ components and the magnitude of $p_{c.m.}$ are shown.  Following Ref.~\cite{Cohen:2018gzh} the $z$-axis points along the $\hat{p}_{miss}$ direction and the $q$-vector lays in the x-z plane.    
The right panel y-axis scale correspond to the measured data counts and the calculations are individually area normalized to the data. The left panel y-axis scale is arbitrary.
Since the T+SCX calculation does not change the shape of the momentum distribution, it is almost indistinguishable from the PWIA calculation. 
	}
\label{Fig:Pcm}
\end{center}
\end{figure*}

\begin{figure}[t]
\begin{center}
\includegraphics[height=5cm, width=\columnwidth]{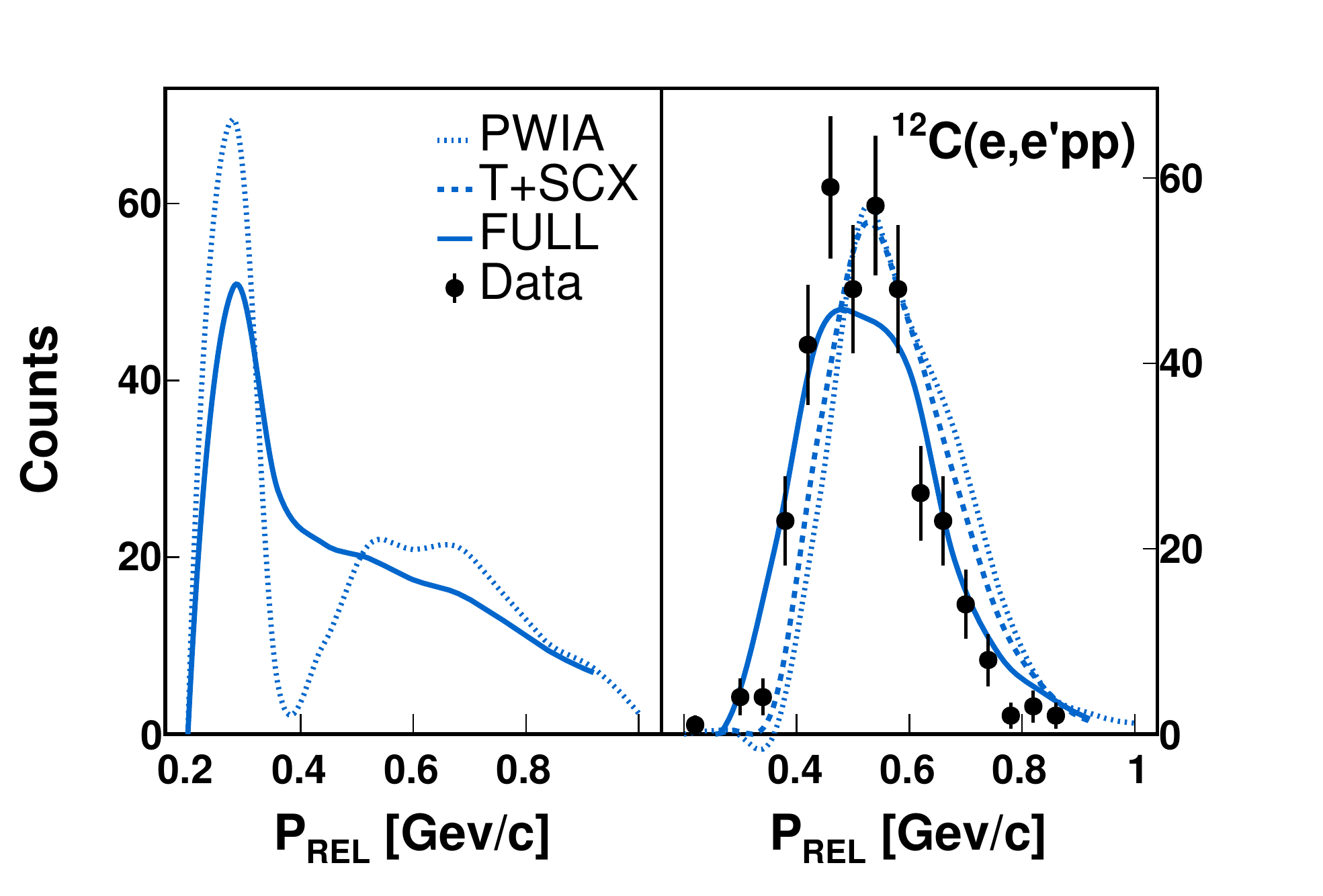}
	\caption{
		Same as Fig.~\ref{Fig:Pcm} for the inferred SRC pair relative momentum in the $^{12}$C$(e,e'pp)$ reaction for scattering off nucleons in SRC pairs.
	}
\label{Fig:Prel}
\end{center}
\end{figure}

We applied the event selection cuts previously used to select kinematics with
minimal sensitivity to FSI effects~\cite{schmidt20}:
\begin{itemize}
\item $x_B > 1.2$,
\item $\theta_{pq} < 25^\circ$,
\item $0.62 < |{\bf p_N}|/|{\bf q}| < 0.92$,
\item $0.4 < |{\bf p_{miss}}| < 1.0$ GeV/c,
\item $m_{miss}$ $\leq$ 1.1 GeV,
\item $|{\bf p_{recoil}}| > 0.35$ GeV/c,
\end{itemize}
where $\theta_{pq}$ is the angle between the leading (knocked-out) proton and the momentum transfer, ${\bf p}_N$ is the leading proton momentum, ${\bf p}_{miss} = {\bf p}_N - {\bf q}$ is the missing momentum, $m_{miss} = \sqrt{(2m + \omega - E_N)^2 - ({\bf p}_N - {\bf q})^2}$ is the ``missing mass'' of the recoil nucleon assuming breakup of a stationary two-nucleon pair, and (if applicable) ${\bf p}_{recoil}$ is the momentum of the detected second proton.  This corresponds to the low-$\omega$ side of the QE peak ($\omega < Q^2/2m$) where both nucleons are emitted from the nucleus in the general direction of the momentum transfer.

In order to test the effectiveness of these cuts in reducing the effects of  FSI, we compared the for the predictions of the different calculations for various kinematical distributions before and after the application of the SRC selection cuts. We only consider events with kinematics that can be detected in CLAS.
We study both $^{12}$C$(e,e'p)$ and $^{12}$C$(e,e'pp)$ events samples which differ only in the requirement of detecting a recoil proton in the final state. 
As the transparency and SCX probabilities used in the T+SCX calculation were only determined for the limited kinematics of events passing our event selection cuts we do not consider that calculation when studying the full phase-space of GCF events detectable by CLAS.

We first compared 'effective transparencies 'in the T+SCX and Full calculations. We calculated the integrated number of events within the CLAS acceptance that pass the SRC event selection cuts for the PWIA, T+SCX, and Full calculations.  
These effective transparencies are defined as the ratio of the number of T+SCX or Full events to PWIA events (i.e., the number of events including FSI effects divided by the number of events prior to FSI). 
The resulting T+SCX and Full effective transparencies are 0.61 and 0.58 for the $^{12}$C$(e,e'p)$ reaction and 0.83 and 0.73 for the $^{12}$C$(e,e'pp)$ reaction. 
The larger ratio observed for the $^{12}$C$(e,e'pp)$ reaction is due to its increased sensitivity to SCX reactions following an interaction with an $np$-SRC pair as the effective transparency factors we report on account for both attenuation and SCX.
Overall the fact that the PWIA/Full and PWIA/T+SCX ratios agree to $5-15\%$ is very encouraging.

Next we looked at how FSI distorts measured momentum distributions.
Figs.~\ref{Fig:p_lead} and~\ref{Fig:pp_lead_recoil} show the distribution of the measured protons for $^{12}$C$(e,e'p)$ and $^{12}$C$(e,e'pp)$ events respectively.  The left panel shows all SRC breakup events detectable by CLAS while the right panel shows the subset of events passing our SRC selection cuts. FSI have a significant impact on the full measurable phase-space, but a much smaller effect on the selected SRC events. FSIs have a larger effect on the detected recoil nucleon momentum than on the leading nucleon momentum.   The full GENIE calculation agrees better with the data than the PWIA and/or T+SCX calculations. 
This is most evident for the recoil proton momentum distribution at low momenta where FSI distort the measured distribution.

We next examined the impact of FSI on the measured missing-momentum distribution, see Fig.~\ref{Fig:Pmiss}. Many studies use $\bf{p}_{miss}$ as a proxy for the ground state nucleon momentum, since  in the absence of FSI they are equal.   The SRC selection cuts still strongly suppress FSI effects, especially for the $^{12}$C$(e,e'p)$ channel.  For the $^{12}$C$(e,e'pp)$ channel, the SCX reactions have a larger effect, but the previous T+SCX approach and the new GENIE transport approaches agree.  This is expected as the predominance of $np$-SRCs amplifies the effects of $(n,p)$ charge exchange contributions to the measured $^{12}$C$(e,e'pp)$ events.

Figure~\ref{Fig:pp_p} shows the ${p}_{miss}$ dependence of the $^{12}$C$(e,e'pp)$/$^{12}$C$(e,e'p)$ event yield ratio. The raise of the measured ratio with  ${p}_{miss}$ was interpreted by Ref.~\cite{schmidt20} as evidence for a transition from a Tensor dominated $NN$ interaction at moderate high-momentum (which suppress $pp$-SRCs) to a scalar repulsive core at high-momentum (where $pp$-SRC contribution is enhanced). Here we see that accounting for full FSI via the GENIE transport model does not have much impact on this observable, which bolsters the findings of Ref.~\cite{schmidt20}.

Last we examine in Fig.~\ref{Fig:Pcm} and ~\ref{Fig:Prel} the impact of FSI on the pair c.m. and relative momenta respectively, defined as: $\mathbf{p}_{cm} = \mathbf{p}_{miss} + \mathbf{p}_{recoil}$ and $\mathbf{p}_{rel.} = (\mathbf{p}_{miss} - \mathbf{p}_{recoil})/2$.
The pair relative momentum distribution should be directly sensitive to the two-nucleon interaction at short-distance \cite{Weiss:2015mba,Weiss:2016obx,Cruz-Torres:2019fum}. 
The pair c.m. motion is a measure of the interaction between the SRC pairs and the residual $A-2$ nuclear system.

\begin{figure*}[t]
\begin{center}
\includegraphics[height=4.3cm, width=2\columnwidth]{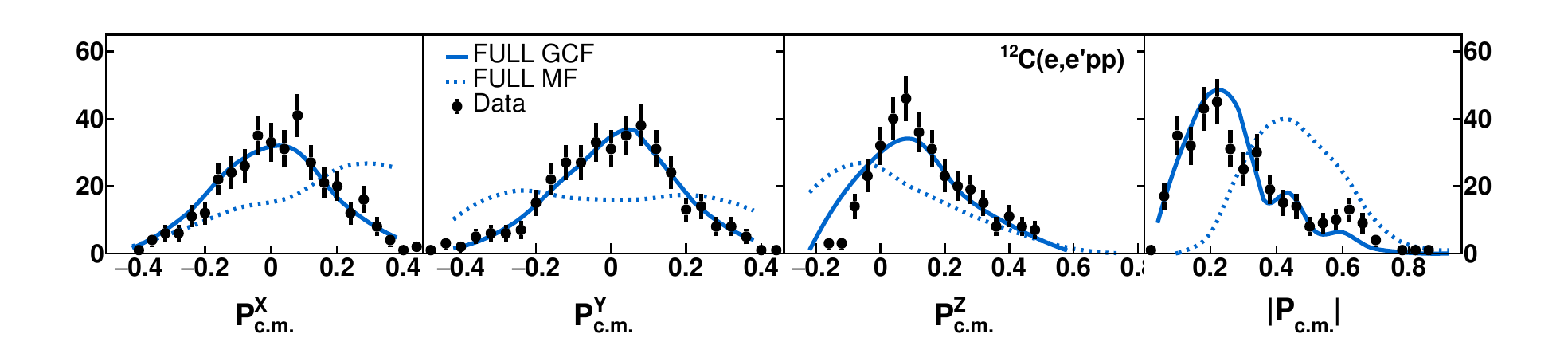}
	\caption{
		Same as Fig.~\ref{Fig:Pcm}, comparing calculations for scattering off MF nucleons and nucleons in SRC pairs as modeled by the GCF.
	}
\label{Fig:MF_Pcm}
\end{center}
\end{figure*}

\begin{figure}[t]
\begin{center}
\includegraphics[height=4.8cm, width=\columnwidth]{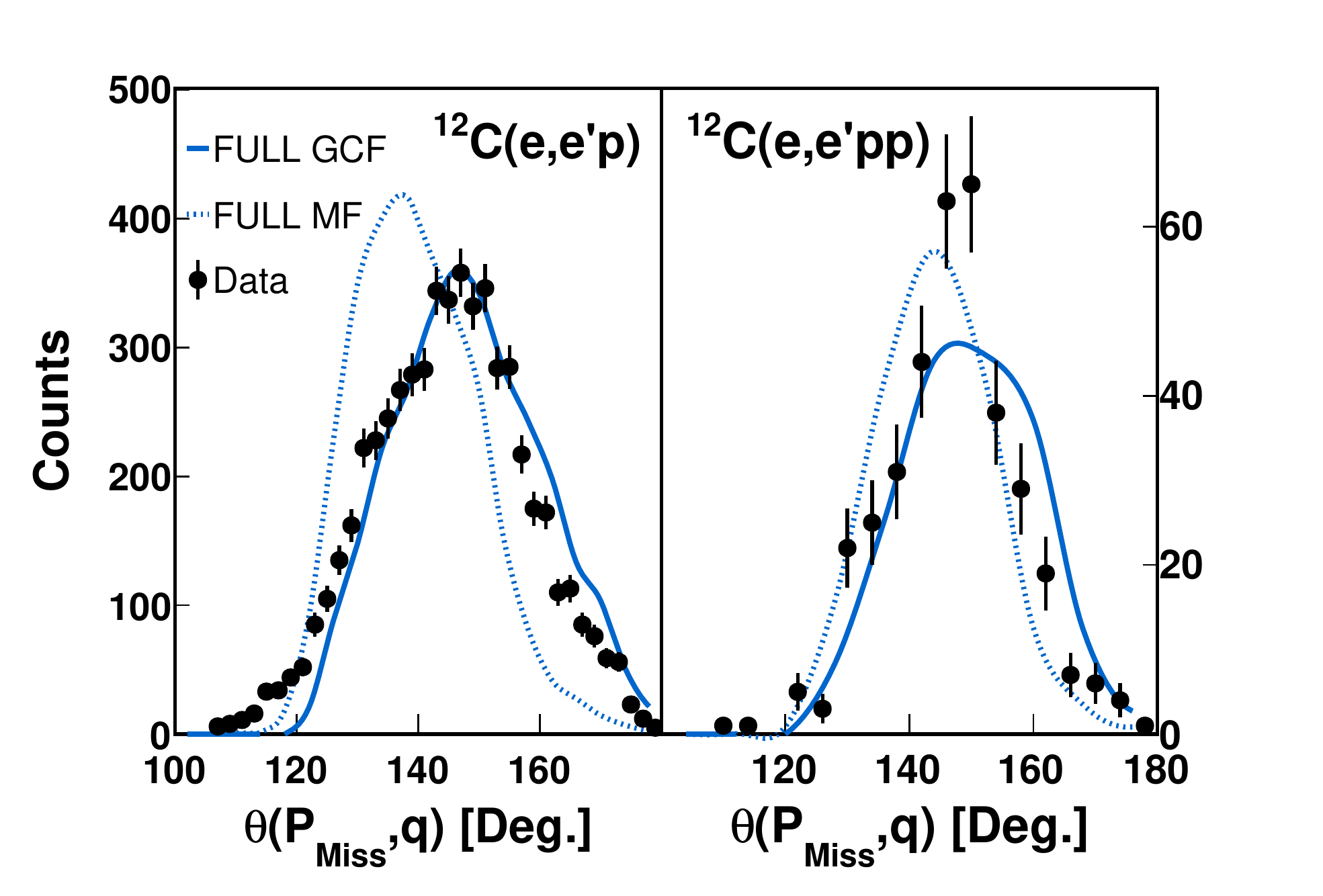}
	\caption{
		Distribution of the angle between the missing-momentum ($\bf{p}_{miss}$) and momentum transfer ($\bf{q}$) vectors for the $^{12}$C$(e,e'p)$ (left) and the $^{12}$C$(e,e'pp)$ (right) reactions.
		Data is from Ref.~\cite{schmidt20} and the calculations and shown for scattering off mean-field (MF) nucleons (dashed) and nucleons in SRC, pairs as modeled by the GCF (solid).
		The y-axis scale correspond to the measured data counts and the calculations are individually area normalized to the data.
		}
\label{Fig:MF_Pmq}
\end{center}
\end{figure}

\begin{figure}[t]
\begin{center}
\includegraphics[height=5cm, width=\columnwidth]{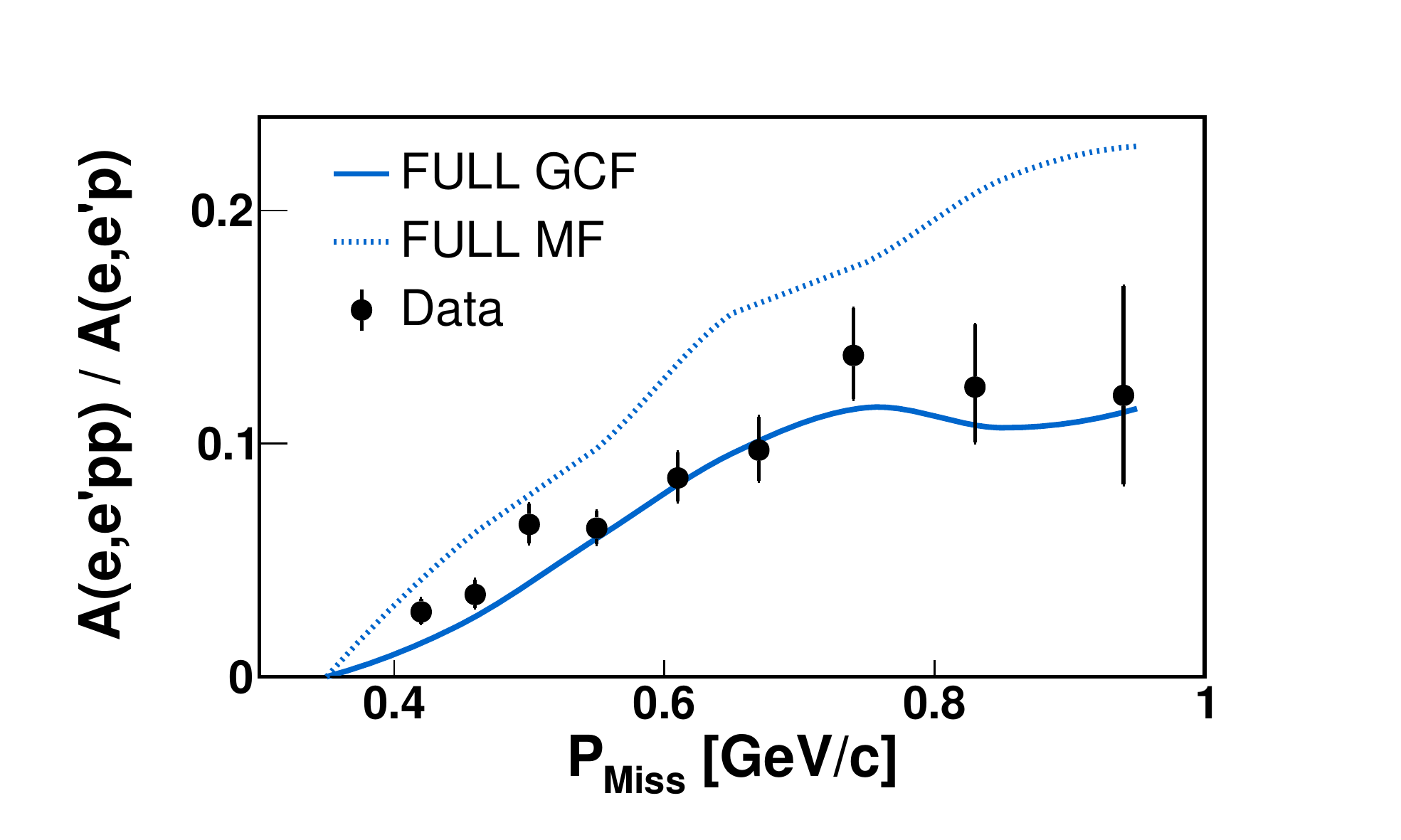}
	\caption{
		Same as Fig.~\ref{Fig:pp_p}, comparing calculations for scattering off MF nucleons and nucleons in SRC pairs as modeled by the GCF.
	}
\label{Fig:MF_pp_p}
\end{center}
\end{figure}

We see that FSI significantly distorts the $\vert \mathbf{p}_{c.m.}\vert$ distribution, both before and after the application of the SRC event selection cuts.  
To gain further insight to this we examined the distributions of the $x, y$ and $z$ components of $\mathbf{p}_{c.m.}$, using coordinate system of Ref.~\cite{Cohen:2018gzh} where the z-axis points along the $\hat{p}_{miss}$ direction and the $q$-vector lays in the x-z plane.
In this case we find that FSI have only a small impact on the $p_{c.m.}$ distributions in the transverse direction, but a large impact in the longitudinal direction.
This strongly supports  the $(e,e'pp)$ data analysis of Ref.~\cite{Cohen:2018gzh}, where the width of the SRC pairs c.m. momentum distribution was extracted from the transverse direction only, and the apparent saturation of the distribution width from C to Pb was taken as evidence for the suppression of FSI that would otherwise broaden the distribution more for larger nuclei.

The PWIA $p_{rel}$ distribution in Fig.~\ref{Fig:Prel}  shows the known minimum for $pp$ pairs at about $400$ MeV/c before the event selection cuts.  This minimum is filled in for $np$ pairs by the tensor force.  Thus, SCX processes that transfer $np$-SRC pair scattering events to $(e,e'pp)$ final state events, fill in that minimum.  After application of our event selection cuts, the different calculations show less of a difference, with the full  calculation showing slightly better agreement with the data.

Last we examined the possible contribution of interactions with mean-field (MF) nucleons (i.e. single nucleons with momentum smaller than $k_F$) that following FSI have kinematics that mimic that of interactions with an SRC pair.
Such events can, in principle, the identical distributions to those of SRC events and can therefore presents an irreducible background for SRC studies.
To test this we repeated the analysis described above for running GENIE using a momentum distribution that follows a local Fermi-Gas model.
At the PWIA level no event passes the SRC selection cuts but following FSI some MF events pass our event selection cuts.

Not surprisingly, many kinematical distributions look very similar to those resulted from the GCF calculation. However, several key distributions show clear differences, which we focus on here.
Fig.~\ref{Fig:MF_Pcm} shows the c.m. momentum distribution measured via the $^{12}$C$(e,e'pp)$ reaction channel, comparing data with GCF and Mean-field calculations.
Both calculations include full eGENIE FSI and are shown after the application of the SRC events selection cuts.
As can be seen, the MF calculation results in a distribution that is  significantly broader than that observed for both the data and GCF calculation.

Figure~\ref{Fig:MF_Pmq} shows the distribution of the angle between $\bf{p}_{miss}$ and $\bf{q}$, which is sensitive to FSI effects.
We shown both data and GCF and Mean-field calculations both both the $^{12}$C$(e,e'p)$ and $^{12}$C$(e,e'pp)$ reactions.
As can be seen, for the $^{12}$C$(e,e'p)$ reaction the MF distribution peaks at $\sim 135^\circ$, in contrast with the data and GCF distributions that peak at $\sim 150^\circ$. From these distributions one cannot exclude that some of the smaller-angle events might contain some MF contamination.
For the $^{12}$C$(e,e'pp)$ reaction the situation is less clear as the two calculations result in somewhat similar distributions that the low statistics of the data does not allow distinguishing between.

Last, in Fig.~\ref{Fig:MF_pp_p} we note a striking difference in the ${p}_{miss}$ dependence of the $^{12}$C$(e,e'pp)$ / $^{12}$C$(e,e'p)$ ratio where the MF calculation has a much higher slope than both the data and the GCF calculation.

In summary, SRC studies using high-energy electron-induced nucleon knockout reactions are highly informative and, at the same time, can be very sensitive to kinematical distortion effects due to FSI. Current high-energy scattering measurements are carried out at highly selective $x_B > 1$ anti-parallel kinematics where reaction theory studies based on Eikonal and Glauber approximations show suppression of such distortion for a wide range of observables. However, these calculations cannot be done either for the exact kinematics studied experimentally or for heavier nuclei. They also struggle modeling FSI effects on the recoil nucleon in two-nucleon knockout reaction which has momentum between 300 and 800 MeV/c and is below the applicability range of Glauber theory.  It is thus useful to study FSI effects via additional, complementary, theoretical models. 

By implementing the GCF model in the GENIE lepton-nucleus interaction simulation framework we are able to use its transport calculations to understand the phase-space available for FSI and thereby study its potential impact on observables under specific kinematical conditions and event selection cuts. We find that the kinematical conditions developed in the past decade for SRC studies using exclusive and semi-inclusive reactions significantly reduce the potential impact of FSI on the general phase-space one could use to study SRCs. Residual FSI distortions are still visible, especially for the lower-momentum recoil nucleon distribution. This distortion does not impact the missing-momentum distributions typically used to study the $NN$ interaction and has minimal impact on relevant observables such as the pair c.m. motion when looking in the transverse direction to the missing-momentum.

Last, we note that the implementation of the GCF in eGENIE will enable future studies of neutrino-nucleus reactions involving SRC pairs

\begin{acknowledgments}
We thank Mark Strikman and Steve Dytman for insightful discussions.
This work was supported by the U.S. Department of Energy, Office of Science, Office of Nuclear Physics under Award Numbers DE-FG02-94ER40818, DE-SC0020240, DE-FG02-96ER-40960, DE-FG02-93ER40771. This manuscript has been authored by Fermi Research Alliance, LLC under Contract No. DE-AC02-07CH11359 with the U.S. Department of Energy, Office of Science, Office of High Energy Physics.
\end{acknowledgments}

\bibliography{../../../references.bib}

\end{document}